\documentclass[fleqn,usenatbib]{mnras}

\usepackage[T1]{fontenc}
\usepackage{ae,aecompl} 

\usepackage{graphicx}   
\usepackage{amsmath}    
\usepackage{amssymb}    

\newcommand{\kms}{km\,s$^{-1}$\,}

\newcommand{\Msunpc}{$M_{\odot}/pc^2$\,}

\newcommand{\Ha}{H${\alpha}$ }

\newcommand{\de}{$^\circ$ }

\newcommand{\Msunyr}{$M_{\odot}\, yr^{-1}\,$}
\newcommand{\Msunyrkpc}{$M_{\odot}\, yr^{-1}\, kpc^{-2}\,$}
\newcommand{\aFe}{[$\alpha$/Fe]~}

\newcommand{\Nf}{132~}
\newcommand{\Nc}{375~}
\newcommand{\Nmis}{16~}

\title[SF-driven outflows in Face-on galaxies]
{SDSS IV MaNGA - Star-Formation Driven Biconical Outflows in Face-On Galaxies}

\author[Bizyaev et al.]{
D. Bizyaev$^{1,2}$\thanks{E-mail: dmbiz@apo.nmsu.edu},
Yan-Mei Chen$^{3,4}$,
Yong Shi$^{3,4}$,
Namrata Roy$^{5}$, \\
\newauthor Rogerio Riffel$^{6,8}$,
Rogemar A. Riffel$^{7,8}$,
Jos\'e G. Fern\'andez-Trincado$^{9}$, \\
$^{1}$Apache Point Observatory and New Mexico State University, Sunspot, NM, 88349, USA\\
$^{2}$Sternberg Astronomical Institute, Moscow State University, Universitetskiy prosp. 13, Moscow, 119234\\
$^{3}$School of Astronomy and Space Science, Nanjing University, Nanjing 210093, China \\
$^{4}$Key Laboratory of Modern Astronomy and Astrophysics (Nanjing University), Ministry of Education, Nanjing 210093, China\\
$^{5}$ Department of Astronomy and Astrophysics, University of California, 1156 High Street, Santa Cruz, CA 95064 \\
$^{6}$ Departamento de Astronomia, Instituto de F\'\i sica, Universidade Federal do
Rio Grande do Sul, CP 15051, 91501-970, Porto Alegre, RS, Brazil \\
$^{7}$ Departamento de F\'\i sica, Centro de Ci\^encias Naturais e Exatas,
Universidade Federal de Santa Maria, 97105-900, Santa Maria, RS, Brazil \\
$^{8}$ Laborat\'orio Interinstitucional de e-Astronomia - LIneA, Rua Gal. Jos\'e Cristino 77, Rio de Janeiro, RJ
- 20921-400, Brazil \\
$^{9}$ Instituto de Astronom\'ia, Universidad Cat\'olica del Norte, Av. Angamos 0610, Antofagasta, Chile\\
}
 
\begin{document}
 
\date{Accepted XXX. Received YYY; in original form ZZZ}

\pubyear{2021}

\label{firstpage}
\pagerange{\pageref{firstpage}--\pageref{lastpage}}
\maketitle

\begin{abstract}
We find 132 face-on and low inclination galaxies with central star formation 
driven biconical gas outflows (FSFB)
in the SDSS MaNGA (Mapping Nearby Galaxies at APO) 
survey. The FSFB galaxies show either double peaked or broadened
emission line profiles at their centres.
The peak and maximum outflow velocities are 58 and 212 \kms, respectively. 
The gas velocity dispersion reveals a mild dependence
on the central star formation surface density compatible with models of 
gas dispersion powered by the Jeans instability in gas clumps or by gas turbulence dissipation.
We estimate the gas outflow rate and conclude that the central gas depletion time
does not depend on galactic mass. In turn, the ratio of the gas outflow rate to the 
gas consumption rate by the star formation is low in massive galaxies and high
in low-mass objects, while the star formation is a more rapid process of the 
gas consumption. 
We compare properties of the FSFB galaxies with a control sample of 375 comparison galaxies
and find that the FSFB objects have high central concentration of star formation, 
and also younger central stellar population with respect to their periphery.
We analysed the environment of the galaxies and 
identified nearby satellites and elements of low surface brightness structure. 
We see that many tidal-enhanced features that can be assigned to early and intermediate
stages of galactic interaction are much more frequent in the FSFB galaxies with respect to 
the comparison sample. 
We conclude that the gas should be replenished via the accretion from 
small satellites.
\end{abstract}

\begin{keywords}
galaxies: kinematics and dynamics, galaxies: ISM, galaxies: interactions,
galaxies: star formation, galaxies: structure, ISM: jets and outflows
\end{keywords}

\section{Introduction}

Gas outflows from central regions of galaxies
can be powered not only by active galactic nuclei (AGN) 
but also by star formation bursts \citep{strickland07,superwinds0,superwinds,heckman15}. 
Multi-phase gas outflows driven by star formation bursts in galaxies without AGNs
are common in both nearby \citep{veilleux05,chen10} and distant \citep{rubin14,davies18} galaxies.
While in most galaxies the outflows eject gas in all directions \citep{superwinds}, some 
objects, including nearby star burst galaxies M~82 and NGC~253, show 
biconical star formation driven superwinds that start from galactic centres
\citep{heckman90,lehnert99}.
Numerical simulations predict that central biconical outflows are common in galaxies
\citep{ttmt98,fielding17,schneider18}. Studies of the centrally localized outflows 
provide a foundation of understanding properties of star forming outflows in
general case. 

The gas ejection speed is a critical parameter necessary
for better understanding the processes of gas circulation driven by superwinds and
for determining main properties of gas motion that affects the lifecycle of gas in galaxies
\citep{veilleux05,zhang18}. The speed ranges from hundred \citep{shopbell98,veilleux05} 
to thousand \kms \citep{heckman00}, which leads to the gas loss rate up to 10-20 \Msunyr
\citep{heckman02,chisholm15}. While galaxies observed at high inclination angles
allow us to detect the gas outflows easily \citep{B19}, their geometry 
does not favor the gas speed measurements. In contrast, 
objects observed at close to face-on inclination angles would allow us
to assess the outflow speed directly. 

Large integral field spectroscopic surveys of last years allow us to 
create large samples of unique galaxies, including the objects with 
central, star formation driven outflows.
In this paper we apply the criterion for selecting galaxies with star-formation 
driven biconical outflows derived by \citet{B19} to galaxies
with low inclination angles found in the sample of MaNGA survey objects from \citet{manga}.
The special orientation of galactic planes
makes the visual selection of these objects difficult, but instead it simplifies 
kinematic studies of such objects. At the same time, elements of galactic structure 
can be identified with better certainty in the low-inclined galaxies with
respect to the galaxies with high inclination studied by \citet{B19}, hereafter B19.

\section{The Sample of Star Formation Driven Bicones from MaNGA observations}

In this paper we use a sample of galaxies observed with the Integral
Field Unit (IFU) survey MaNGA (Mapping Nearby Galaxies at APO), see
\citet{manga,drory15,law16}. The survey was conducted at a 2.5-m telescope \citep{gunn06}
in the frames of the Sloan Digital Sky Survey \citep{york00,blanton17,DR17}.
MaNGA targeted a large sample of galaxies uniformly distributed in stellar
mass \citep{wake17}, and achieved a few percent precision of its flux calibration
\citep{yan16a,yan16b}, two dozen \kms precision in the emission line velocity dispersion estimation
\citep{law21}, and a kiloparsec-scale spatial resolution in mapping galaxies
at the survey's median redshift of 0.03 \citep{yan16a}. 
MaNGA has released its complete sample of 10,050 galaxies in the frames 
of the data release SDSS DR17 \citep{DR17}. The released data products include
maps of emission line fluxes and radial velocities in gas and stellar population
\citep{westfall22} which continuously cover most parts of the galaxies. 

We select low inclination galaxies from the whole MaNGA sample
based on photometric estimates of the inclination angles from the parent
catalogue NSA \citep{nsa}. For the galaxies with the inclination angle 
below 45 degrees, we apply a criterion for selecting 
galaxies with star formation driven bicones using the "figure of merit" F
derived by \citet{B19} as follows:
\begin{equation}
F \,=\, \log [\Sigma^{1/2}_{SFR} Z_c / \sigma_c] - 2\log R_e    ~~~~~,
\end{equation}
where the star formation rate surface density $\Sigma_{SFR}$ is measured at the centre 
from the \Ha luminosity, $Z_c = \tau_V 40*10^{(O/H)}$ at the centre, $\tau_V$ is the optical depth 
proportional to the extinction $A_V$ estimated from the Balmer decrement via 
fluxes in emission lines, $\sigma_c = 1.65 (0.33 v_c -2)$ , $v_c$ is the maximum 
rotation velocity and $R_e$ is the effective radius of the galaxies.
We select the galaxies with F > -5.0.
Additionally, we require that the central gas velocity dispersion exceeds
the stellar velocity dispersion. 
We select 326 galaxies with low inclination and high values of the figure of merit.
As it follows from \citet{B19}, the "figure of merit" is based on a
combination of metallicity, extinction, star formation rate density, central 
velocity dispersion, and the compactness of the central galactic region.
The $F$ is a single parameter that allows us to distinguish between the 
galaxies with star formation driven biconical outflows and regular galaxies. 
Figure 9 in \citet{B19} shows that the former galaxies prefer to have
high $F$ values, which indicates high star formation surface density,
high metallicity and large compactness at their central regions. 
Objects with $F > $ -5.0 have a high probability to harbor the
star formation driven bicones. Significantly lower values of the $F$
allow us to select regular galaxies for comparison purposes, see below.


After browsing SDSS images \citep{DR17} of the candidates we removed 
the cases of galaxy mergers, which includes all cases of objects
with overlapping bodies significantly distorted by interactions. 
The BPT diagrams \citep{bpt} of remained
objects were inspected and only the galaxies with star forming centres
were left in the sample. Only the central 3x3 spaxel areas were considered. 

\begin{figure}
\includegraphics[width=\columnwidth]{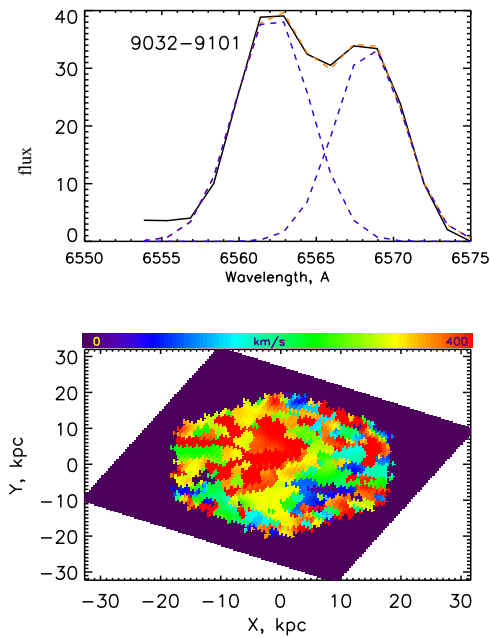}
\caption{
Top: An example of double picked line profile for one of our galaxies with biconical outflow
and our Gaussian fitting. The solid black line designates the observing line profile,
the blue dashed lines show two gaussians separately, while the red dashed 
line demonstrates the sum model profile.
Bottom: the distribution of the gas velocity dispersion in a FSFB galaxy.
\label{fig1}}
\end{figure}

As the next step, we analyzed \Ha emission line profiles in the central spaxel 
and found that most of selected
face-on galaxies with central star formation driven biconical gas outflows
(FSFB)
have widened lines or show clear double peak profiles. 
We fit the line profiles with double Gaussians. We assume that the blue Gaussian line is 
associated with approaching side of gas outflow, while the red Gaussian component
corresponds to its receding side. We find that the central line profiles of 116 galaxies out of 248 
can be successfully fitted with a single Gaussuan, while \Nf galaxies (53\%) require two Gaussians. 
An example of 
a Gaussian fitting is shown in Figure~\ref{fig1} in the top panel. 
The bottom panel shows that 
the gas velocity dispersion is also high at the centres of the galaxies with wide central
emission lines. 

\begin{figure}
\includegraphics[width=\columnwidth]{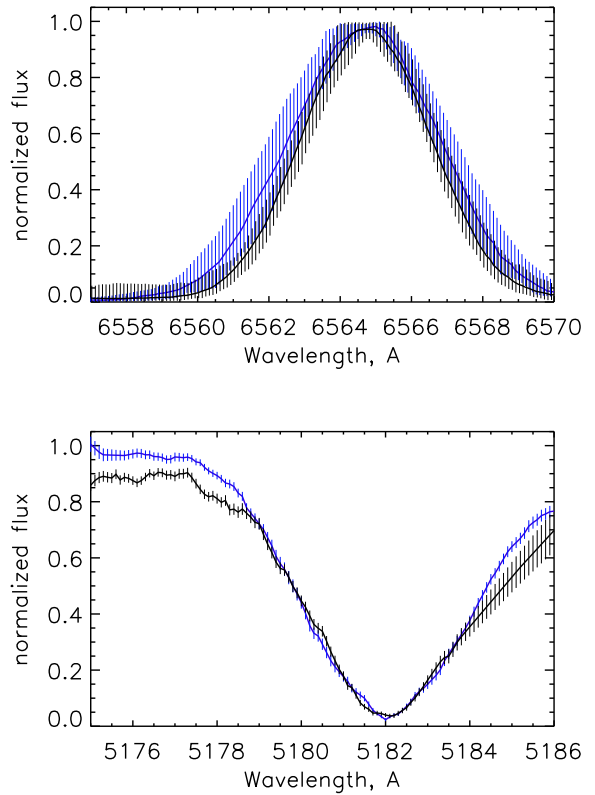}
\caption{
Top: the median co-added profiles (solid curves) of 
the \Ha line for the FSFB (blue) and comparison (black) galaxies. 
The error bars show 10 and 90 percentile curves for both samples. 
Bottom: the same for the Mgb absorption feature. 
It is seen that the FSFB sample has wider average \Ha line profile, while the Mgb absorption feature
has the same width for both samples of galaxies. 
\label{fig1a}}
\end{figure}

To ensure that the FSFB sample has wider emission lines with respect
to the control sample, we aim to compare
the central \Ha emission line profile among the two sample, and to compare a prominent
absorption line profile at the same time.
Figure~\ref{fig1a} compares averaged profiles of
\Ha emission line (top panel) and averaged profiles of a prominent absorption Mgb line (bottom) for both 
samples of galaxies - FSFB and control. The line profiles are considered at the central spaxel of the 
galaxies. Note that all line profiles in Figures~\ref{fig1} and \ref{fig1a} are corrected to the 
radial velocity that corresponds to zero stellar velocity, which corresponds to the wavelength 
of \Ha line in vacuum (6564.6 \AA). The selected FSFB galaxies have systematically wider emission 
line profiles, while the absorption line profile does not show significant difference 
between the samples. Note that the combined \Ha line profile for the FSFB sample 
is a combination of double-peaked and widened profiles, so although the combined curve 
doesn't indicate distinctive double-peaked shape, it has a wide, flat top and large
error bars. Being more narrow, the combined profile for the comparison objects
is closer to a single gaussian shape.
It can be seen that the emission line profiles are symmetric with respect to
the zero velocity of the stellar component. This suggests that we observe the biconical
outflows rather than the blue outflow peak and the redder peak corresponding to the centre of 
galaxy.

As a result, we selected \Nf galaxies, which we refer to as "face-on galaxies with star formation driven biconical outflows" or FSFBs. We also inspected velocity fields of emission gas and stars in all of these galaxies
and found \Nmis cases of significant misalignment between stellar and gas kinematics, with position angles difference exceeding 30\de, according to the KINEMETRY package \citep{krajnovic06}.  

We aim to compare the FSFB objects with "regular galaxies", so we make a comparison 
sample of galaxies with similar properties but without the biconical outflows. 
We started with selecting low inclination galaxies same way as described above,
then we reverted the selection criterion and left only objects with $F < -5.3$.
We checked the central \Ha line profiles same way as for the FSFB group and found only 
8 out of 483 objects with widened \Ha profile at their centres, which require two
Gaussians to fit the \Ha line profile. These galaxies were 
removed from the comparison group. Note that all these 8 galaxies are found in objects 
with noticeable interaction with their satellites. We inspected optical SDSS images 
of the comparison galaxies and removed 60 more cases of merging galaxies. 

\begin{figure}
\includegraphics[width=\columnwidth]{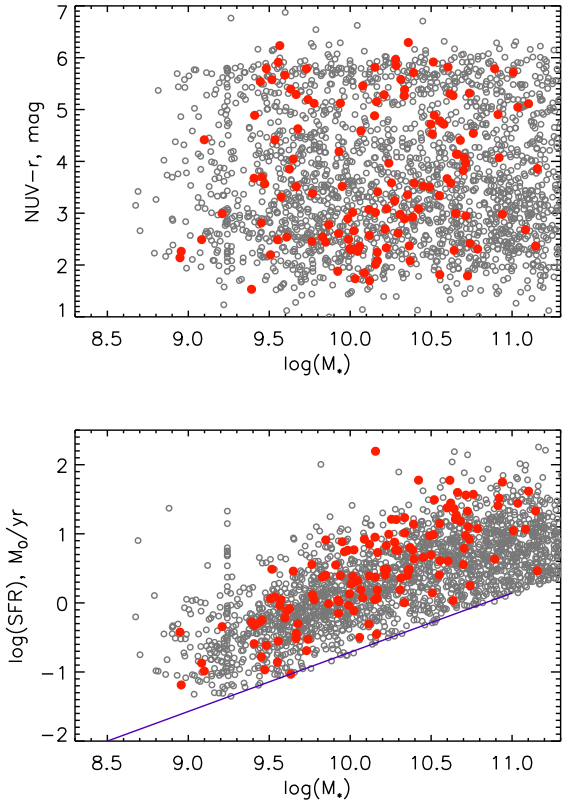}
\caption{
The MaNGA galaxies with biconical outflows (FSFB, red bullets) and without them 
(grey circles, comparison sample) are shown on the SFR - Stellar Mass diagrams. 
Top: the near-ultraviolet - red magnitude difference from the NASA-Sloan Atlas (NSA),
as a proxy for the specific star formation rate. 
Bottom: the stellar mass is taken from the NSA catalog, while the integrated star 
formation rate is estimated via the \Ha fluxes provided by MaNGA. 
The solid line demarcates star forming objects (above) from green valley 
and quiescent galaxies \citep{chang15}.
\label{fig2}}
\end{figure}

When we place the FSFB and comparison groups on the star formation rates (SFR) versus 
stellar mass ($M_*$) diagrams in Figure~\ref{fig2}, we notice that all FSFB galaxies 
are located above the line that demarcates star forming and green valley galaxies,
according to \citet{chang15}. While most of objects in the comparison sample are located above
this demarcation line, too, a small fraction of them lays in the 
"green valley" region. We removed those objects from the comparison sample. 
As a result, the comparison group comprises \Nc galaxies.  
The resulting distribution on the SFR-$M_*$ diagram is demonstrated in Figure~\ref{fig2}.

\begin{figure}
\includegraphics[width=\columnwidth]{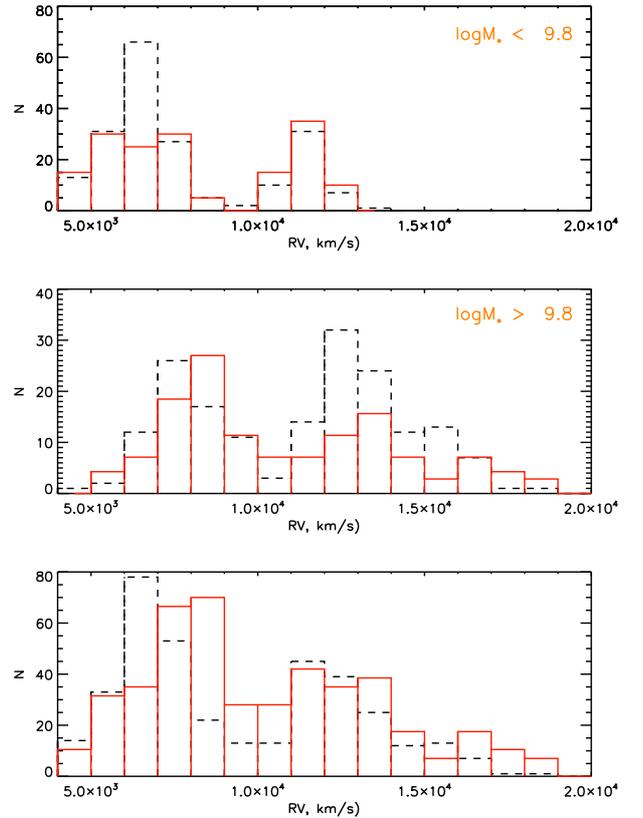}
\caption{
Radial velocities of selected face-on galaxies with SFB (red line) and comparison galaxies 
(black dashed line). The top and middle panels show low- and high-mass galaxies,
respectively, The bottom panels shows all mass groups together. 
\label{fig3}}
\end{figure}

As a consistency check, we plot histogram distributions of radial velocities
for the selected groups of galaxies, see Figure~\ref{fig3}. The histograms
ensure that galaxies in our FSFB and comparison
samples have rather similar distance distributions.

\section{Properties of the Galaxies with Star Formation Driven Bicones}

\subsection{Structural Parameters}

Our selection procedure with equation (1) intentionally biases parameters of FSFB galaxies with respect to the control sample, so we expectantly see systematic difference in the effective radius and metallicity between the two samples.  
Given the face-on orientation of galaxies in our samples, we can study the radial distributions of parameters without significant projection effects, which makes our comparison between the samples more straightforward 
than that for the galaxies with high inclination used by \citet{B19}. 

\begin{figure}
\includegraphics[width=\columnwidth]{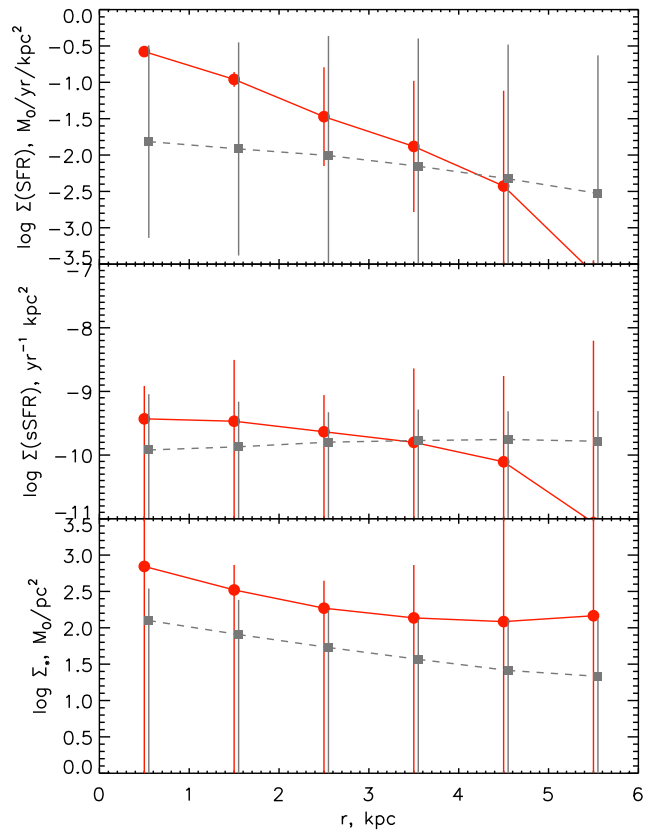}
\caption{
The median values and 1-$\sigma$ error bars for the star formation rate surface density (top), 
specific star formation rate surface density (middle), and stellar surface density (bottom) 
for the FSFB galaxies (red symbols and solid lines) and control sample (grey symbols and dashed lines).
\label{fig4}}
\end{figure}

\begin{figure}
\includegraphics[width=\columnwidth]{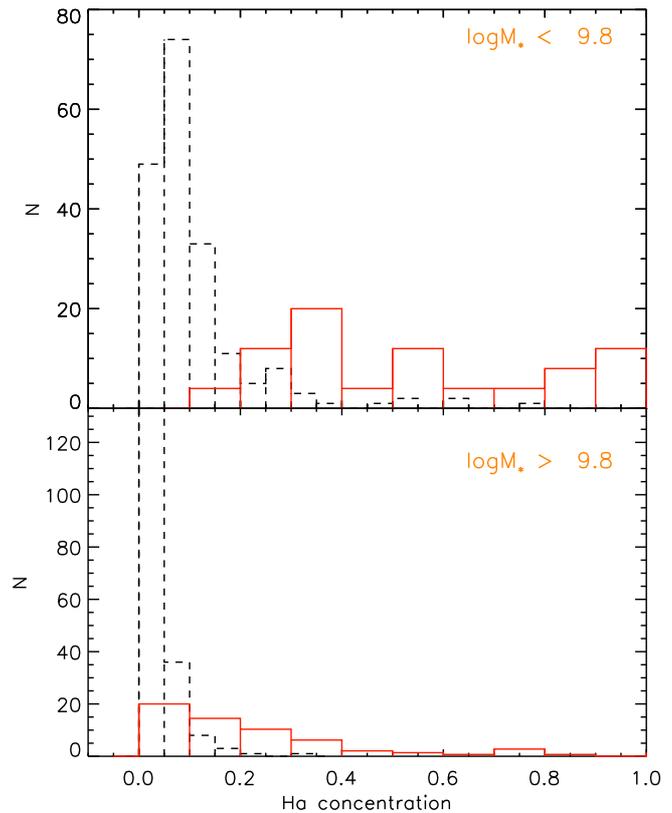}
\caption{
Concentration of \Ha emission in low- (upper panel) and high-mass (lower panel) galaxies. The designation is kept the same as in 
Figure~\ref{fig3}.
\label{fig5}}
\end{figure}

Figure~\ref{fig4} shows how the surface density of star formation, specific star formation
(defined as sSFR = SFR/$M_*$ or as the similar ratio of the surface densities) and stellar 
mass densities change with radius. We stacked values
in annuli regularly spaced along the radius. The bullets and error bars correspond to the 
mean and standard deviation values, respectively. 
Similar to B19,
we see a significantly enhanced central concentration of star formation rate in the FSFB galaxies - 
only in the very central radial bins. The specific star formation rate is not significantly
different between the main and control samples. 
We highlight this difference even more in Figure~\ref{fig5}, where the \Ha concentration
is shown for low- ($\log M_*/M_{\odot} < 9.8$) and high-mass ($\log M_* < 9.8 /M_{\odot}$) galaxies. 
Same as in B19, the \Ha concentration is estimated as the ratio of \Ha luminosity within and outside 
the central 1 kpc circle.
While the fraction of centrally concentration is higher in massive FSFB galaxies with 
respect to the comparison objects, this difference becomes especially prominent 
in the low mass galaxies. 

\subsection{Stellar Population}

\begin{figure}
\includegraphics[width=\columnwidth]{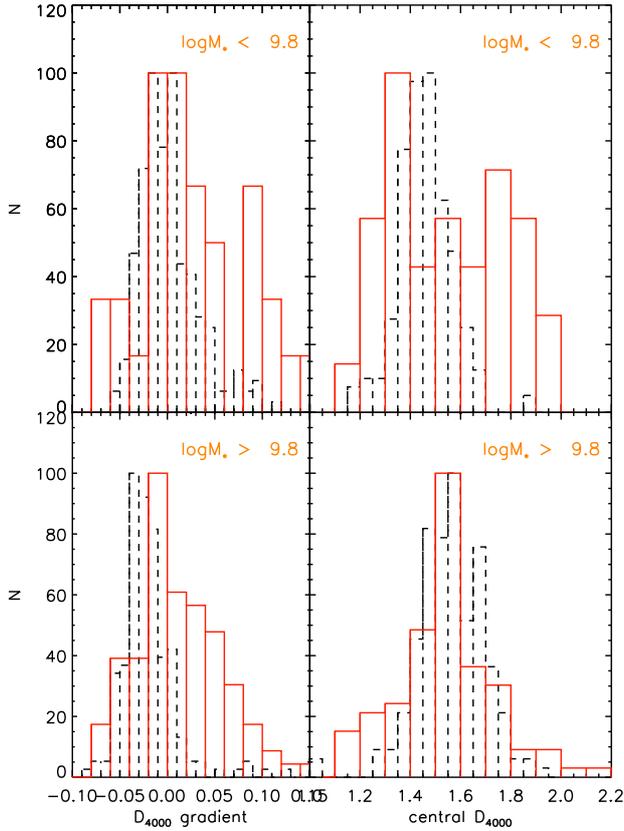}
\caption{
The central values and radial gradient of D(4000) index for the FSFB (solid red lines) 
and control galaxies (dashed black lines) for low- and high- mass galaxies (top and
bottom panels, respectively).
\label{fig6}}
\end{figure}

While properties of stellar population in the galaxies observed at high
inclination angles studied by B19 is affected by projection effects,
face-on orientation allows to consider a  more clear case. 
Figure~\ref{fig6} shows results of a linear fitting of D(4000) radial index
distribution in each galaxy: the index central value and its radial gradient.
Low mass FSFB galaxies have more objects with positive D(4000) gradients than the
comparison galaxies. The central values have more scatter in them, 
and the distribution reveals more FSFB objects with high central 
values of the index. 
High mass FSFB sample have even a higher fraction of galaxies with 
positive radial gradient of D(4000), while their central values are 
the same as in the comparison galaxies. 

According to numerical estimates of the D(4000) index by \citet{kauffmann03},
the main age of stellar population in our massive galaxies is of the order of 1 Gyr 
in both samples. Centres of low massive comparison galaxies are younger, while 
the FSFB galaxies from this mass range are often a few times older.
Both low mass and especially high mass FSFB galaxies often have
significant positive radial gradient of stellar population age.

\begin{figure}
\includegraphics[width=\columnwidth]{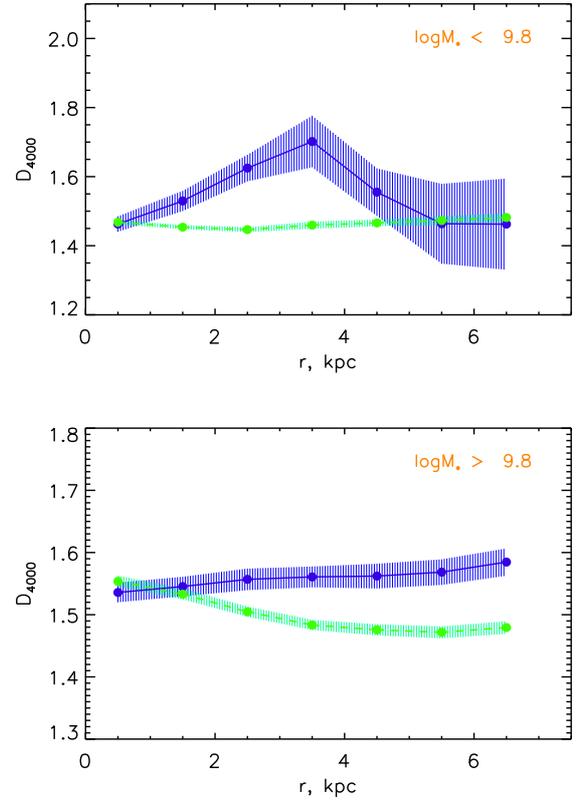}
\caption{
Cumulative radial distributions of mean D(4000) index (curves) and its errors of mean
(shaded areas) for FSFB (blue) and comparison (green) galaxies. The panels 
correspond to the low- (top) and high-mass (bottom) galaxies. 
\label{fig6a}}
\end{figure}

Figure~\ref{fig6a} shows the mean value of D(4000) index and its error of the mean
determined in 1 kpc wide radial annuli evenly distributed along
the radius. While the centres of the galaxies have essentially the same age between
the samples, FSFBs show consistently higher positive gradient with respect to 
the comparison sample. FSFB objects have younger stellar population at their
centres in the comparison with the rest of the galaxy. The regular
galaxies, in contrary, have older centres with respect to their periphery.

\begin{figure}
\includegraphics[width=\columnwidth]{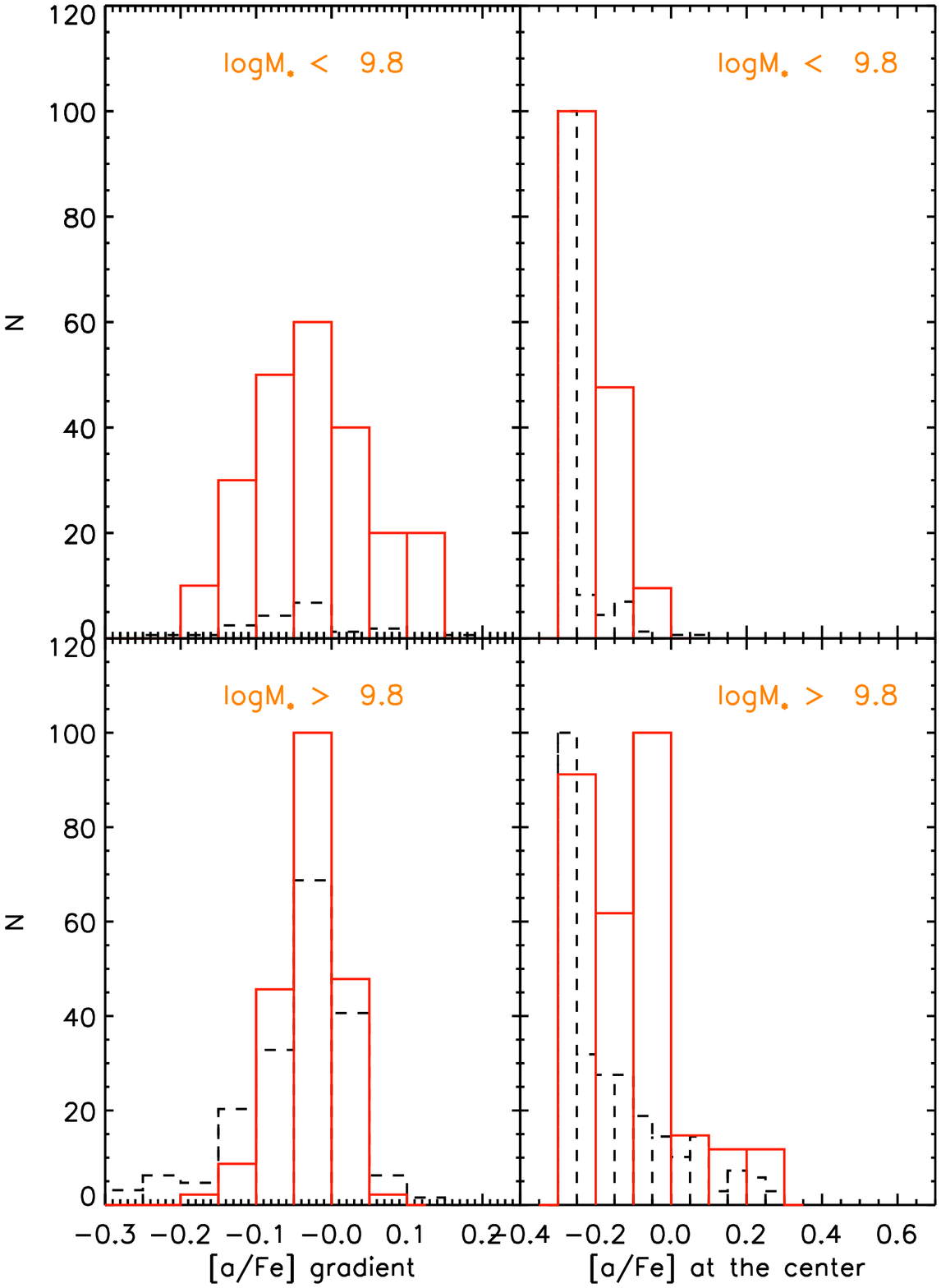}
\caption{
The central values and radial gradient of \aFe index for the FSFB (solid red lines) 
and control galaxies (dashed black lines) for low- and high- mass galaxies (top and
bottom panels, respectively).
\label{fig7}}
\end{figure}

The radial gradient of the \aFe index demonstrated in Figure~\ref{fig7}
does not show significant 
difference between the FSFB and control samples. 
Although the histogram distributions look slightly different
for the low- and high-mass galaxies in the left panels, the 
mean and mode values of the distributions for the FSFB and control
galaxies look similar.
In contrast, 
FSFB galaxies of all masses often show higher \aFe at their centres
than at the periphery. This can be interpreted as significant contribution 
of recent starburst (within the last Gyr) to the formation of galactic centres
of FSFB objects. 

\subsection{Kinematics of Ionized Gas}

While the face-on orientation of galactic midplanes makes difficult their selection,
it simplifies measuring the outflow velocities. 
As it was mentioned in \S2, we analyze \Ha emission line profiles at the central
spaxel. All galaxies in our sample have wide or double profiles that need to be fitted with two 
Gaussian lines. Many galaxies have clearly separated double peaks. Velocities of the 
blue ($v_{blue}$) and red ($v_{red}$)
peaks of fitted Gaussians are assigned to the approaching and receding sides of  
ionized gas outflows. We assume that the mean central outflow speed is equal to 
$V_{out} \,=\, 0.5 (v_{red} - v_{blue})$. We also assume that the FWHM of the Gaussians
$dV_{out}$ indicates the real velocity distribution in the outflows, and that the
maximum gas ejection speed is $V_{max} \,=\, \sqrt{ V_{out}^2 +  dV_{out}^2 }$.

\begin{figure}
\includegraphics[width=\columnwidth]{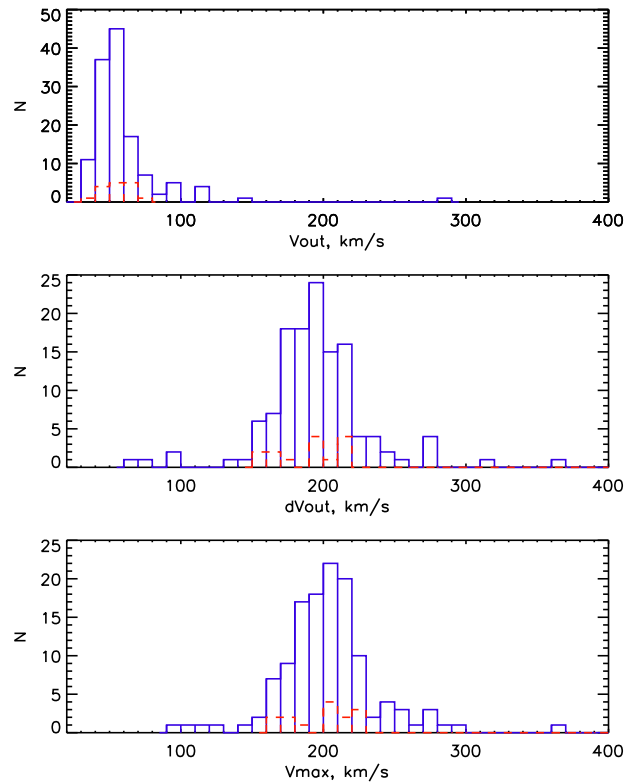}
\caption{
Histogram distributions of gas outflow kinematic characteristics for the FSFB galaxies.
Top: the peak outflow speed in \kms. 
Middle: the FWHM of the outflow component in FSFB in \kms. 
Bottom: maximum outflow speed in \kms, see text. 
The blue line designates all FSFB galaxies, while the red dashed line highlights the galaxies with counter-rotation between their gas and stars. 
\label{fig9}}
\end{figure}

Figure \ref{fig9} shows the histogram distributions of the $V_{out}$, $dV_{out}$ and 
$V_{max}$ in our FSFB galaxies. Out of all galaxies FSFB, 16 objects have counter-rotating
gas and stellar disks. We mark these galaxies with the red dashed line in Figure \ref{fig9}.
It is seen that the subsample of counter-rotators has the same histogram distributions
as the main FSFB sample. 
The mean and median values of the outflowing gas speed are shown in Table~(\ref{tab1}).
Note that our maximum speed is consistent with the gas outflow speed if it were measured 
via the maximum profile width.

\begin{table} 
\begin{tabular}{lcc}
\hline
   &  Mean & Median \\
   &  \kms & \kms \\
\hline
 $V_{out}$ &  58 &  53 \\
$dV_{out}$ & 203 & 195 \\
 $V_{max}$ & 212 & 205 \\
\hline
\end{tabular}
\caption{The mean and median peak outflow speed, FWHM, and maximum
outflow speed for the sample of FSFB galaxies.
\label{tab1} }
\end{table}


\begin{figure}
\includegraphics[width=\columnwidth]{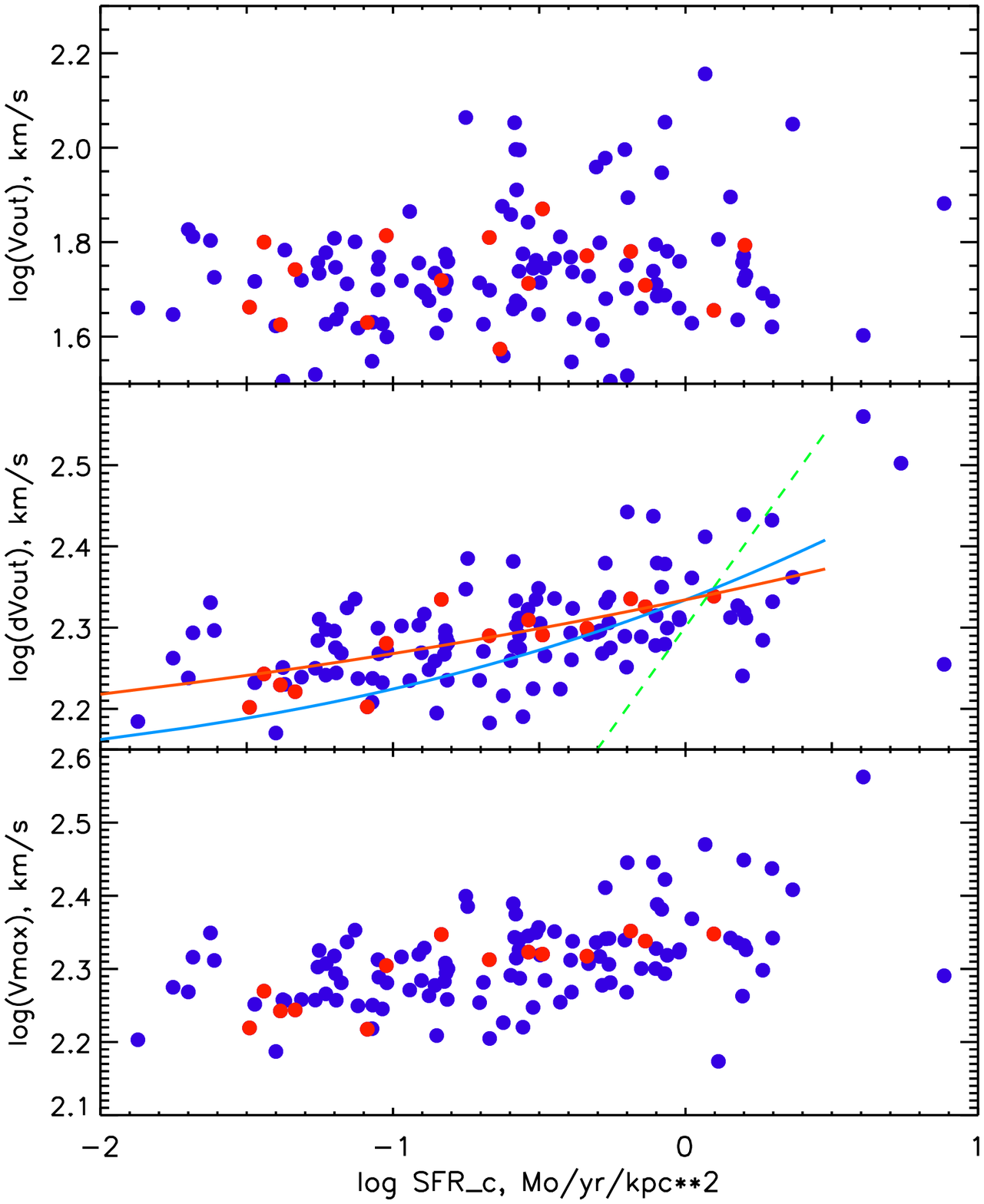}
\caption{
The peak outflow speed (top), its FWHM (middle) and maximum outflow speed (bottom) 
versus the star formation surface density at the centres of FSFB galaxies.
The blue symbols designate all FSFB galaxies, while the red ones mark the galaxies 
with counter-rotation between their gas and stars.
The red, blue, and green curves in the middle panel correspond to 
the gas velocity dispersion caused by the Jeans instability in clumps,
the turbulence dissipation, and the turbulence driven
by gravitational instabilities, respectively (see text).
\label{fig10}}
\end{figure}

\citet{heckman02} found that the star formation outflows occur when the 
star formation surface density exceeds certain limit, e.g. 0.1 \Msunyrkpc. 
Figure \ref{fig10} shows that we have significant number of galaxies in our
sample whose central star formation rate surface density is in the range
from 0.01 to 0.1 \Msunyrkpc. Figure \ref{fig10} demonstrates how the 
peak outflow speed, its FWHM and maximum outflow speed depend on the star
formation surface density at the centres of FSFB galaxies. 

Figure \ref{fig10} shows that the peak gas velocity does not depend significantly on
the central star formation activity $SFR_c$, while the gas velocity dispersion in ejected ionized gas 
is mildly correlated with the $SFR_c$. A linear regression fit to the latter relation has a slope of 0.18. 
This case better corresponds to the case of the gas velocity dispersion caused by
the Jeans instability in clumps, according to \citet{elmegreen07,lehnert09}. 
In this case  $\sigma_{gas} \sim \Sigma_{SFR}^{0.18}$, which is designated with the 
solid red line in the middle panel. 
The small slope of the observing FWHM - $\Sigma_{SFR}$ relation is also in agreement with 
the turbulence dissipation model by \citet{dib06,lehnert09}, 
where $\sigma_{gas} \sim \Sigma_{SFR}^{1/3}$ (shown with the solid blue line).
The green line designates the case of turbulence driven by gravitational instabilities
\citet{lehnert09,krumholz16}, which implies $\sigma_{gas} \sim \Sigma_{SFR}^{0.5}$. 
This slope is too steep for our sample, see the green dashed line in the middle panel.

\begin{figure}
\includegraphics[width=\columnwidth]{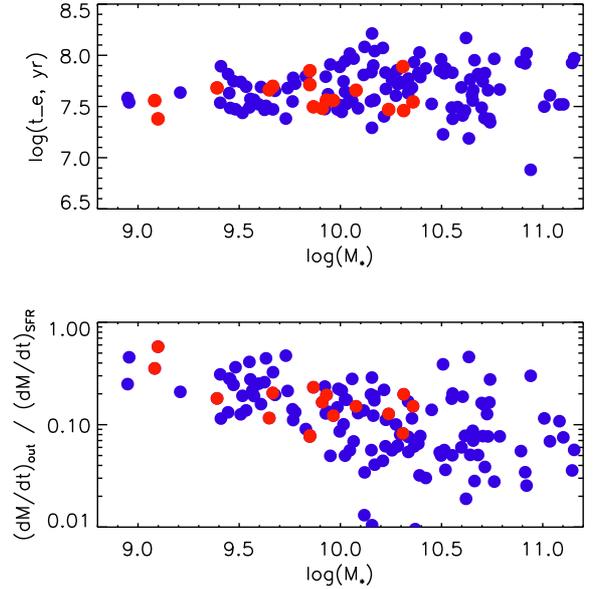}
\caption{Top: the central gas exhaust time for FSFB galaxies
of different stellar mass. Bottom: the ratio of the mass
outflow rate to the star formation rate at the centre of FSFB 
versus the galactic stellar mass.
The blue symbols mark all FSFB galaxies, while the red ones highlight the galaxies with counter-rotation between their gas and stars.
\label{fig11}}
\end{figure}

It is hard to measure the volume density of the ejected gas directly because of projection effects.
At the same time, we can easily measure the size of the \Ha-emitting nucleus and guess
its gas content. The size can be estimated directly as the radius where the \Ha emission
surface density drops e times from its central value. While we cannot measure the
gas surface density directly, we can use the Kennicutt-Schmidt relation between the 
gas surface density and the star formation rate from \citet{kennicutt98} which, in turn, 
is a function of the \Ha luminosity density \citep{martin01}. 
In the assumption of spherical 
geometry of the region with the starburst and biconical outflow, we estimate the gas mass $M_g$,
its volume density and the mass outflow rate (dM/dt)$_{out}$. Figure~\ref{fig11} shows the 
gas outflow exhaust time $t_e$ in the upper panel, 
which is defined from the central gas mass and the gas outflow rate as $t_e \,=\, M_g / (dM/dt)_{out}$.
There is no clear trend of (dM/dt)$_{out}$ and $t_e$ with the galactic mass. The median value 
of $t_e$ is 46 Myr. The bottom panel in Figure~\ref{fig11} demonstrates the ratio
of the gas outflow rate to the star formation rate. The ratio is less than one in all galaxies, which is consistent
with similar estimates for other types of galaxies \citep[e.g.][]{concas22}.
This ratio is systematically low
in massive galaxies, while in low-mass objects it reaches 50\%.

\subsection{Environment of the Galaxies}

Here we study difference in environment between the two samples. 
Two types of environmental features are considered. 
First, we visually inspect images of the galactic vicinity at distances 
comparable to their size using optical SDSS\footnote{http://skyserver.sdss.org}
and DES Legacy\footnote{http://legacysurvey.org/pubs/} images. 

\citet{atkinson13} identified a variety of low surface brightness structures
that can be found around regular galaxies. 
Adopting technique from \citet{atkinson13}, we identify a set of structural elements 
for each galaxy in both FSFB and comparison samples. 
The elements include nearby small satellites (ss), nearby large satellites (ls),
low surface brightness loops (lo), shells (sh), arcs (ar) around the galaxies, 
tails (ta), bridges (br), noticeable asymmetry
of outer isophotes (ao), irregular dust lanes crossing bodies of galaxies (du),
inner (ir) and outer (or) rings, significant inner asymmetry (ia), and wrecked shape (wr). 
We also notice the shape disturbance caused by interaction (di), membership of a tight
group (tg), and an interaction with a companion (in). 
We distinguish the latter
from merging, when two or more objects are forming single stellar structure.

In addition to looking for structural peculiarities in and around the galaxies, 
we widen the field of consideration and count the number of small or large 
satellites in 15 arcmin fields centered at the galaxies. We record the number 
of the following elements: small satellite (SS - very small galaxies around, 
not taken into account in the narrow fields considered above); 
large satellite (LS - large enough galaxies 
with recognized internal structure, but still less in size than the main galaxy); 
even larger satellite (GS - comparable to the main galaxy or slightly larger);
and huge satellite (HS - much larger companion in the field).
Note that some of the SS and HS can result from the objects projection, and not from 
the real proximity in space.  

\begin{figure}
\includegraphics[width=\columnwidth]{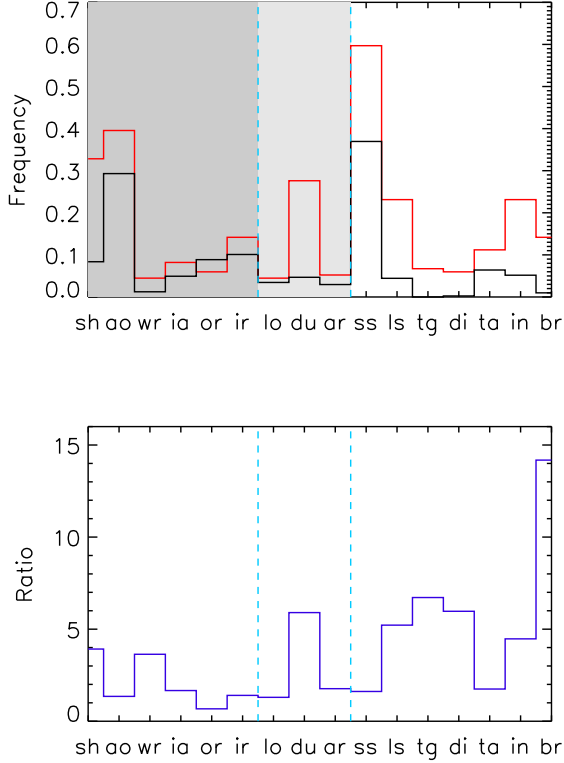}
\caption{
Upper panel: the lines show the frequency of environmental features (shown on the X-axis)
among the entire sample of FSFB (red) and comparison (black) objects.
The areas are shaded differently according to the age of the features (old-intermediate-young
from the left to the right, see text).
Lower panel: The ratio of FSFB to the control galaxies in each bin from the upper panel. 
The vertical blue dashed lines demarcate the features of different age, same as in the upper panel.
\label{fig12}}
\end{figure}

The upper panel in Figure~\ref{fig12} demonstrates the frequency of the structure 
elements defined above in the FSFB (red line) and control (black line) samples.
The lower panel in Figure~\ref{fig12} shows the same information as a 
division of the FSFB frequency by that of the comparison sample, determined for 
each environmental element. Note that 
some structural elements are rarely found in the comparison galaxies, so the results
of the division shown in the lower panel are not reliable numerically, albeit significant.
We notice significantly enhanced fraction of nearby large satellites, shells, 
peculiar dust lanes, disturbed or wrecked shape, signs of interaction, including bridges, 
and association with tight groups in the FSFB objects with respect to the comparison galaxies. 

\begin{figure}
\includegraphics[width=\columnwidth]{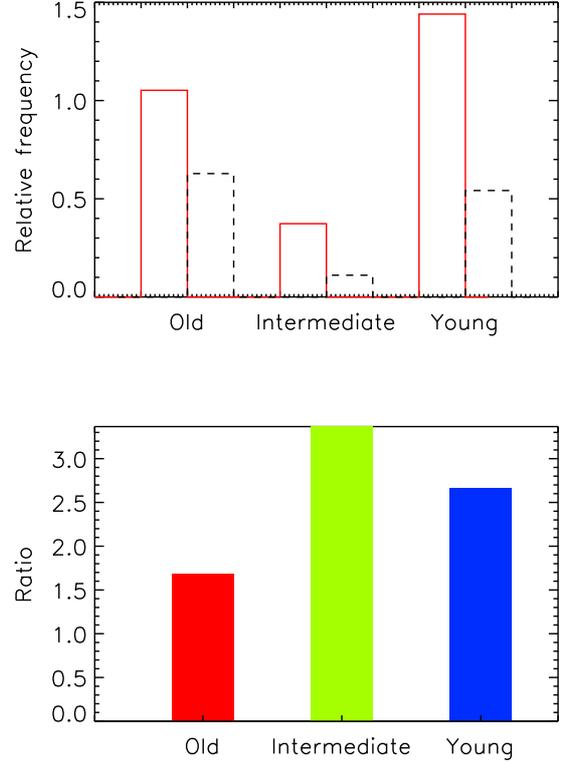}
\caption{
Top: The relative frequency of features from Figure~\ref{fig12} combined by
their age of the interaction for the FSFB 
(red solid line) and comparison (black dashed line)
samples. 
Bottom: the ratio of the frequencies from the top panel.
\label{fig13}}
\end{figure}

We subdivide the environmental features around the galaxies by three classes
based on their interaction stage: young (ss, ls, tg, di, ta, in, br), 
intermediate (lo, du, ar) and old (sh, ao, wr, ia, ir, or).
When combined by their age of interaction, the structural elements also demonstrate 
difference between the samples of galaxies. Figure~\ref{fig13} shows a relative frequency
of the elements in the upper panel for the FSFB (red) and control sample (black).
Their ratio is shown in the lower panel, which reveals the most significant 
difference between the samples of objects with the young and intermediate age features. 

\begin{figure}
\includegraphics[width=\columnwidth]{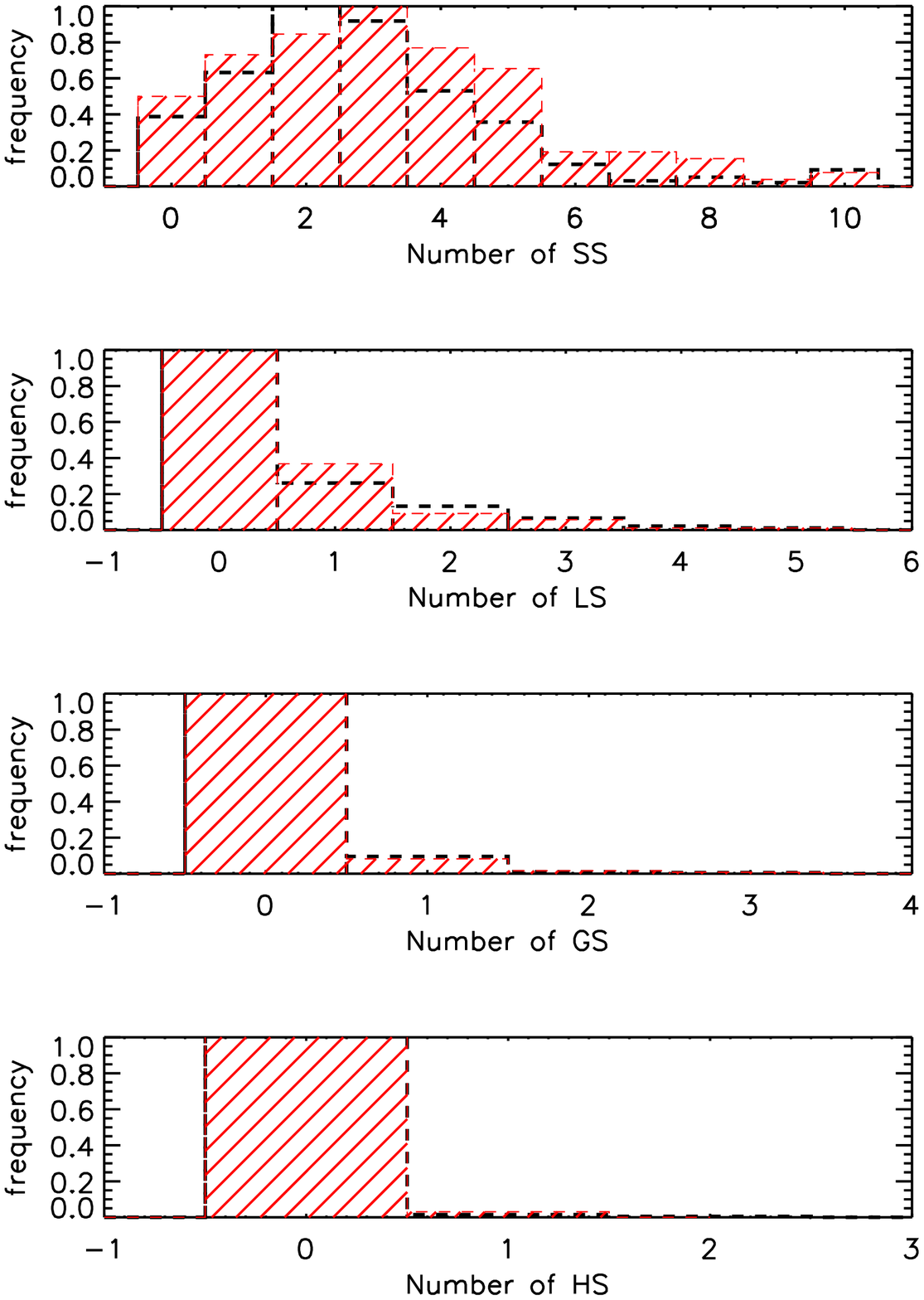}
\caption{Histogram distributions of the number frequency of satellites around the 
FSFB (red lines) and comparison (black dashed lines) galaxies separated
by their size (small to large, from top to bottom).
\label{fig14}}
\end{figure}

Figure \ref{fig14} demonstrates the frequency of satellites between the two groups of galaxies. 
While the large neighbors, like GS, HS and also LS, show no difference between the sample,
the number of smallest satellites is enhanced around the FSFB objects.

\section{Discussion}

Studying MaNGA galaxies observed at low inclination angle allows us to verify
and complete conclusions made in the previous work by B19. Employing advantages
of low inclination, we can measure the gas outflow speed directly. The surface
density of the star formation rate also can be studied without any projection 
effects. Figures~\ref{fig4} and \ref{fig5} confirm that the face-on galaxies
with star formation driven bicones have high central concentration of
\Ha emission and star formation, especially in galaxies with low mass. 
Radial distributions of the stellar population age look special for the 
FSFB galaxies. Figures~\ref{fig6}, \ref{fig6a} and \ref{fig7}
suggest that while galaxies in the comparison sample have older
centres and younger peripheries, the FSFB objects in contrary have
younger central regions. The D(4000) and \aFe distribution in them suggest
that the bulk stellar population at the centres was formed within last 
Gyr. 

Figure~\ref{fig10} demonstrates that although most of the bicones
have their central star formation rate surface density above 0.1 \Msunpc,
a noticeable fraction of them indicates lower star formation rate density
between 0.01 and 0.1 \Msunpc.

Our FSFB galaxies have moderate peak outflow velocity estimated
directly either from a double peaked profile or from broadened 
emission line at the centre, see Figure~\ref{fig9}. Figure~\ref{fig10}
compares the velocity and its dispersion in gas with 
central star formation surface density, and helps make conclusion that the Jeans
instability in star formation clumps and also the turbulence 
dissipation in gas drive the gas turbulence in the central regions of galaxies
with biconical gas outflows and provide a small slope of the $\sigma_c \div 
\Sigma_{SFR}$ relation there. 

Our estimate of the gas outflow rate and of the central gas exhaust time 
shows no difference between low- and high-mass galaxies. The exhaust time scale
is rather short, 46 Myr on average. 
The gas outflow plays a more significant role in the gas 
depletion process in low mass FSFB galaxies with respect to the 
star formation. In contract, star formation is the dominating
process that regulates the gas consumption at the centres of massive galaxies, where the gas outflow
contributes only a small fraction to the gas depletion process. 
At the same time, centres of all our FSFB galaxies convert more gas to stars than
eject to the circumgalactic medium. 

Comparison between the FSFB and control samples of galaxies allows us
to conclude that some features coincide with the central star formation 
driven biconical outflows. Thus, the frequency of shells, inner rings and
unusual dust pattern is much higher in the FSFB objects, as well as
such signs of ongoing or recent interaction as disturbed or wrecked
shape, bridges between galaxies or presence of nearby large neighbors, as well as
membership in tight galactic groups. The presence of four or more small satellites
is also a feature more often seen around the FSFB objects. 

All features of FSFB galaxies mentioned above suggest that 
the galaxies experience starbursts at their centres that 
enhance the fraction of young stellar population.
The small- and large-scale gas turbulence caused by the starburst
helps eject metal-enriched gas with the peak speed of the 
order of 60 \kms, and maximum speed over 200 \kms. At the same
time, the star formation consumes much more gas than massive galaxies
eject, while low-mass galaxies lose their gas via its outflow
more intensively with respect to its depletion by star formation. 
Interaction with 
nearby satellites, ongoing and past, as well as some low surface brightness
environmental structures reveal evidences of connection between the central gas outflow and
interaction with environment in the FSFB galaxies. In a combination with 
younger centres in the FSFB galaxies, we can assume that minor interactions
are responsible for driving gas to the centres of galaxies, where
it fuels intensive starbursts. 

Given the much shorter time scale
of the gas exhausting at the centre (dozens of Myr) in the comparison with 
the age of central stellar population (a few
hundred Myrs), we can conclude that the gas at the centre needs to be replenished. 
Minor interactions with gas-rich satellites should be a source of fresh gas.
Gas supply from the circumgalactic medium should also play some role, but we 
don't clearly see sources of this kind of accretion in our data.

We notice that FSFB galaxies with significant misalignment between gas and stars
show the same features, trends and numerical estimates as the galaxies with 
regular gas and stellar rotation. Since the counter-rotation is a direct evidence
of past interactions of galaxies with their satellites, we are enabled to 
conclude that such interaction might take place in the past of all other
FSFB galaxies.

\section{Summary}

We select \Nf galaxies with low inclination and with biconical gas outflow from the centres driven by 
star formation processes (FSFB) from the final MaNGA data release. 
Our selection procedure is based on a combination of general galactic parameters
that can be determined from photometry and spectroscopy, not necessarily panoramic.
It means that the selection procedure can be run for a much wider samples available from the 
modern sky surveys.
We form a sample of comparison sample of \Nc galaxies with similar characteristics.
All FSFB galaxies show either double peaked or broadened emission lines at their centres.  
We estimate that the mean peak outflow gas velocity is 58 \kms, and the mean maximum
gas velocity is 212 \kms. The slope of the gas velocity dispersion versus the star formation 
rate surface density at the centre is small, what suggests that the gas dispersion is 
powered by the Jeans instability in gas clumps or by the gas turbulence dissipation. 

We conclude that the gas loss rate is significant at the centres of small galaxies
with respect to the gas depletion via star formation processes, while in large 
galaxies the latter is the principal process of the central gas depletion. In combination
with a median gas depletion scale of 46 Myr typical for galaxies of all masses
and the mean stellar population age of a few hundred Myrs, we conclude that the central 
gas should be refilled, most probably from the accretion of small satellites. The latter is
confirmed by enhanced frequency of young and intermediate age features and high number
of small satellites that we see around the FSFB objects, with respect to 
the comparison galaxies. 

Our FSFB sample has \Nmis
galaxies with significant misalignment between kinematics of their gas and stars.
These galaxies show the same trends, features and numerical values as all other
FSFB objects with regular rotation, which also suggests that past interactions with 
small satellites played significant role in supplying gas to the central regions of
galaxies and in the consequent formation of star forming driven biconical gas outflows.


\section*{Acknowledgements}
The authors would like to thank the anonymous referee for constructive
comments that improved the paper. 
Y. C. acknowledges support from the National Key R\&D Program of China (No. 
2017YFA0402700), the National Natural Science Foundation of China (NSFC
grants 11573013, 11733002)
Y.S.  acknowledge support
from the National Key R\&D Program of China (No.  2018YFA0404502) and the
National Natural Science Foundation of China (NSFC grants 11733002 and
11773013). 
The project is partly supported by RSCF grant 22-12-00080.
RR thanks to Conselho Nacional de Desenvolvimento Cient\'{i}fico e Tecnol\'ogico  
( CNPq, Proj. 311223/2020-6,  304927/2017-1 and  400352/2016-8), Funda\c{c}\~ao de 
amparo \`{a} pesquisa do Rio Grande do Sul (FAPERGS, Proj. 16/2551-0000251-7 and 19/1750-2), 
Coordena\c{c}\~ao de Aperfei\c{c}oamento de Pessoal de N\'{i}vel Superior (CAPES, Proj. 0001).
RAR acknowledges financial support from Conselho Nacional de Desenvolvimento Cient\'ifico e 
Tecnol\'ogico (CNPq -- 302280/2019-7) and Funda\c c\~ao de Amparo \`a pesquisa do 
Estado do Rio Grande do Sul (FAPERGS -- 21/2551-0002018-0).
J.G.F-T gratefully acknowledges the grant support provided by Proyecto Fondecyt Iniciaci\'on 
No. 11220340, and also from ANID Concurso de Fomento a la Vinculaci\'on Internacional para 
Instituciones de Investigaci\'on Regionales (Modalidad corta duraci\'on) Proyecto No. FOVI210020, 
and from the Joint Committee ESO-Government of Chile 2021 (ORP 023/2021).


SDSS-IV acknowledges support and resources from the Center for
High-Performance Computing at the University of Utah.  The SDSS web site
is www.sdss.org.

SDSS-IV is managed by the Astrophysical Research Consortium for the
Participating Institutions of the SDSS Collaboration including the
Brazilian Participation Group, the Carnegie Institution for Science,
Carnegie Mellon University, the Chilean Participation Group, the French
Participation Group, Harvard-Smithsonian Center for Astrophysics,
Instituto de Astrof\'isica de Canarias, The Johns Hopkins University,
Kavli Institute for the Physics and Mathematics of the Universe (IPMU) /
University of Tokyo, Lawrence Berkeley National Laboratory, Leibniz
Institut f\"ur Astrophysik Potsdam (AIP), Max-Planck-Institut f\"ur
Astronomie (MPIA Heidelberg), Max-Planck-Institut f\"ur Astrophysik (MPA
Garching), Max-Planck-Institut f\"ur Extraterrestrische Physik (MPE),
National Astronomical Observatory of China, New Mexico State University,
New York University, University of Notre Dame, Observatrio Nacional /
MCTI, The Ohio State University, Pennsylvania State University, Shanghai
Astronomical Observatory, United Kingdom Participation Group, Universidad
Nacional Aut\'onoma de M\'exico, University of Arizona, University of
Colorado Boulder, University of Oxford, University of Portsmouth,
University of Utah, University of Virginia, University of Washington,
University of Wisconsin, Vanderbilt University, and Yale University.

\section*{DATA AVAILABILITY}
This work makes use of SDSS/MaNGA project data publicly available at
https://www.sdss.org/dr17/data\_access/.

{}

\bsp    
\label{lastpage}

\end{document}